\documentclass[aps,prl,reprint,groupedaddress,preprintnumbers,showpacs]{revtex4-1}
\usepackage{graphicx}
\usepackage{dcolumn}
\usepackage{bm}
\usepackage{amsmath} 
\usepackage{amsfonts} 
\usepackage{amssymb}
\usepackage{color}

\begin{document}

\preprint{KEK-TH-1381}

\title{Non-Equilibrium Fluctuations of Black Hole Horizons}

\author{Satoshi Iso}
\email[E-mail address: ]{satoshi.iso@kek.jp}
\author{Susumu Okazawa}
\email[E-mail address: ]{okazawas@post.kek.jp}
\author{Sen Zhang}
\email[E-mail address: ]{zhangsen@post.kek.jp}

\affiliation{KEK Theory Center, Institute of Particle and Nuclear Studies,\\
High Energy Accelerator Research Organization(KEK)\\
and\\
The Graduate University for Advanced Studies (SOKENDAI),\\
Oho 1-1, Tsukuba, Ibaraki 305-0801, Japan}

\date{\today}

\begin{abstract}
We investigate non-equilibrium nature of  fluctuations of black hole horizons
by applying the 
fluctuation theorems and the Jarzynski equality
developed in the non-equilibrium statistical physics.
These theorems applied to space-times with black hole horizons lead to 
the generalized second law of thermodynamics.
It is also suggested that the second law should be violated 
microscopically so as to satisfy the Jarzynski equality.
\end{abstract}


\maketitle

{\it Introduction.} --- There are striking similarities between
space-time with horizons and thermodynamic systems. The analogies
have been extensively investigated, especially, in the black hole thermodynamics.
A black hole behaves as a blackbody with the Hawking temperature 
$T_H= \hbar \kappa/2 \pi$ \cite{Hawking:1974sw}, and energy $\Delta E$ 
flowing into the black hole can be identified as the entropy increase
$T_H \Delta S_{BH}$ of the black hole. Here, $\kappa$ is the surface gravity 
at the horizon and the entropy of the black hole $S_{BH}$ is 
proportional to the area of the horizon $A$ as
$S_{BH}=A/4G\hbar$ in the Einstein-Hilbert theory of gravity.
The thermal behavior is essentially quantum mechanical.
Furthermore, such a thermal behavior is not restricted 
to globally-defined horizons like an
event horizon of a black hole, but also applicable to local horizons
such as the Rindler horizon of a uniformly accelerated observer. 
The point is emphasized in e.g. \cite{Jacobson:2003wv}. 
At the classical level, this is related to an interplay of 
a local change of the horizon area and the Einstein equation.
In particular,
T. Jacobson pointed out  \cite{Jacobson:1995ab}
that the Einstein equation is derivable from 
a thermodynamic relation for the local Rindler horizon.
At the quantum level, the local notion of horizon entropy has 
more fundamental meaning, since it gives transition rates
of area-changing irreversible processes of black holes,
and will be related to the quantum statistical
nature of the space-time. 
If there exists a fundamental quantum structure of space-time
behind the horizon thermodynamics, how can we probe such
microscopic states?  
\vspace{0.5em}
\\
In order to tackle this question, let us go back into history
at the early twenties century. 
Einstein proposed several methods to determine the Avogadro constant $N_A$.
In particular he investigated the dynamics of the Brownian motion in
\cite{Einstein:1905}. 
His aim was giving a proof of the reality of atoms. 
If we knew nothing about the atoms, how could we count the number of 
atoms contained in one mole gas?
His idea was to measure fluctuations of physical quantities around
thermal equilibrium.
In thermodynamics, 
the unit of energy per kelvin per mole is given by the gas constant $R$,
and any thermodynamic  observable in equilibrium is
expressed in terms of $R$. 
As far as one measures thermal averages in equilibrium,
it is impossible to disentangle the macroscopic unit of energy $R$ 
into the Avogadro constant $N_A$ and the Boltzmann constant $k_B$. 
Einstein's insight was in paying attention to fluctuations around 
thermal equilibrium.  
A thermal quantity  fluctuates because of
the microscopic structures behind the thermodynamics. 
He reversed 
the entropy formula of Boltzmann $S=k_B \log W$ 
as $W=\exp (S/k_B)$, and obtained the mean squared values 
of fluctuations in terms of the microscopic unit of energy $k_B$
and various response functions \cite{Einstein:1910}. For example, 
the entropy fluctuation at constant pressure
is given by $\langle (\Delta S)^2 \rangle =k_B C_p$ where 
$C_p$ is the heat capacity at constant pressure. 
This is now understood as a fluctuation-dissipation theorem,
which relates the fluctuation around the equilibrium to a 
response function. 
\vspace{0.5em}
\\
After Einstein, many important works have been performed to
understand out-of-equilibrium properties of fluctuations.
One of the most notable developments will be the non-equilibrium 
fluctuation theorem first discovered in numerical simulations 
\cite{Evans:1993}. 
There are several variations of the theorem.
An adequate one in the present context 
is the Crooks fluctuation theorem \cite{Crooks:1998}.
Assume that the system is initially 
in thermal equilibrium at inverse temperature
$\beta$ with an external parameter $\lambda^F(0)$.
Then one changes the external parameter $\lambda^\text{F}(t)$  as a 
function of time from $t=0$ to $t=T$. The system evolves
according to the Hamiltonian dynamics. The procedure of
changing the parameter corresponds, 
for example, to a process of moving a piston and it needs not to be quasi-static.
In changing the external parameter,
the system becomes out-of-equilibrium. 
For each microscopic state, one measures the amount of work
exerted on the system, and takes an ensemble average
over the initial density matrix.
Define $\rho^\text{F}(W)$ as a probability that the exerted work is given by $W$
under the change of parameter $\lambda^\text{F}(t)$.
$\lambda^F(t)$ is called a forward protocol.
We also consider a reversed protocol in which we change the
external parameter in a reversed way as $\lambda^\text{R}(t)\equiv \lambda^\text{F}(T-t)$
from $t=0$ to $t=T$. 
The system is assumed to be initially in thermal equilibrium
at the same temperature, but with a different external parameter 
$\lambda^\text{R}(0)=\lambda^\text{F}(T)$. Define 
$\rho^\text{R}(-W)$ as a probability that the work is given by $-W$
in the reversed protocol. 
The Crooks fluctuation theorem states that the ratio
of these two probabilities
is given in terms of the work and the difference of free energies $F(\lambda)$
between the two equilibrium states,
\begin{align}
\frac{\rho^{\text{F}}(W)}{\rho^{\text{R}}(-W)}=e^{\beta({W}-\Delta F)},
\end{align}
where we defined $\Delta F= F(\lambda^\text{R}(0)) -F(\lambda^\text{F}(0))$.
From the Crooks fluctuation theorem, the 
Jarzynski equality \cite{Jarzynski:1997} can be obtained,
$\langle e^{-\beta W}\rangle =e^{-\beta \Delta F}$.
Here, the angled bracket stands for the average with the probability $\rho^{\text{F}}(W)$.
It is surprising since the average of exponentiated work in non-equilibrium 
processes in the left hand side is related to the 
difference of equilibrium quantities at the beginnings of the protocols.
These relations imply that
non-equilibrium fluctuations are miraculously arranged so as
to satisfy such very nontrivial relations.
By using the Jensen's inequality, 
$\langle \exp(x) \rangle \ge \exp (\langle x \rangle)$,
the Jarzynski relation is reduced to 
$\langle ( W - \Delta F) \rangle \ge 0$,
which implies the second law of thermodynamics.
Though the average of $W-\Delta F$ is always non-negative,
the Jarzynski relation requires that there is a nonzero probability
for the quantity to take a negative value microscopically.
\vspace{0.5em}
\\ 
The purpose of this letter is to apply these celebrated relations of non-equilibrium 
fluctuations to space-times with horizons. 
In particular, we consider non-equilibrium fluctuations of black hole horizons.
\vspace{1em}
\\ 
{\it Transition Rates.} --- We consider a coupled system of a black hole
and  matter. 
The external parameter  $\lambda(t)$  characterizing
the system Hamiltonian, 
which appeared in the fluctuation theorem, 
 can be arbitrarily chosen here such as a height of a potential
for the matter. If the whole system is controlled by a unitary 
time evolution with time-reversal symmetry, 
a transition probability ${\cal W}_{\lambda^\text{F}(t)}({\cal C} \rightarrow {\cal C}')$
from one configuration ${\cal C}$ to another
${\cal C}'$  under a time evolution of the external parameter 
$\lambda^\text{F}(t)$ is equal to a probability ${\cal W}_{\lambda^\text{R}(t)}({\cal C}' \rightarrow {\cal C})$
from ${\cal C}'$ to ${\cal C}$ under
the reversed change of the parameter $\lambda^\text{R}(t)$.
In the presence of a black hole, however, 
the time-reversal symmetry is violated by imposing the 
ingoing boundary condition at the horizon. 
In a black hole space-time, regular coordinates near
the horizon, i.e. Kruskal coordinates $(U,V)$,
are defined by $U = -\kappa^{-1} e^{-\kappa (t-r_*)}$ and
 $V = \kappa^{-1} e^{\kappa (t+r_*)}$. Here, $t$ and $r_*$ are 
the Schwarzschild-time 
and the tortoise coordinates. 
Quantum fields near the horizon are classified into two
types of chiral fields (in a two-dimensional sense on $(t,r_*)$ plane),
one depending on $U$ and the other on $V$. Fields depending on $V$
are ingoing waves falling into the black hole while those depending on $U$
are propagating nearly along the horizon and correspond to outgoing modes.
The regularity at the horizon requires  occupation of outgoing modes 
$\phi(\omega) \sim e^{-i\omega U}$ to vanish at the (future) horizon. 
Namely, we must impose the vacuum condition for the outgoing modes in the
Kruskal coordinates. On the contrary, there is no constraint for the
ingoing modes, and the conditions are
 asymmetric between $U$ and $V$.
The time-reversal transformation $t \rightarrow -t$ exchanges 
the coordinates $U$ and $V$, and the presence of horizon violates
the time-reversal symmetry of the quantum states.
Therefore,
the above transition probabilities are not necessary the same.
\vspace{0.5em}
\\ 
The ratio of the above transition probabilities 
was evaluated by Massar and Parentani \cite{Massar:1999wg}
for Hawking radiation processes.
They have shown that
the transition rates for systems with a black hole horizon 
are governed by changes in the horizon area.
In the present case, the ratio of a transition probability 
between a configuration ${\cal C}$
with black hole area $A$ and another one ${\cal C}'$ with area $A'$
under a fixed value of external parameter $\lambda$
is given by
\begin{align}
\frac{{\cal W}_\lambda({\cal C}(A) \rightarrow {\cal C}'(A'))}{{\cal W}_\lambda({\cal C}'(A') \rightarrow {\cal C}(A))} = \exp (\Delta A/4G\hbar),
\label{ratio}
\end{align}
where the change of area $\Delta A=A'-A$ is assumed to be small. 
In deriving this, 
they used the WKB approximation for the system wave function,
and calculated the transition rates in 
the first Born approximation for the interaction between the detector and 
radiation field.
A similar result was obtained in a different way in \cite{Kraus:1994by}.
If we identify $\Delta A$  
as the energy $\Delta E$ emanated from the black 
hole by $\Delta A/4G\hbar = - \Delta E/T_H$,
it becomes the Boltzmann factor  $\exp(-\Delta E/T_H)$
in the Hawking radiation. 
The ratio eq.(\ref{ratio}) takes into account the back reaction 
of the geometry against emanating radiation quanta. 
If detailed balance is satisfied in the processes, 
the ratio eq.(\ref{ratio}) is identified 
as the ratio of a probability
in the configuration ${\cal C}'(A')$ to that in ${\cal C}(A)$,
and hence consistent with  the entropy of a black hole $S_{BH}=A/4G\hbar.$
\vspace{0.5em}
\\ 
In the proof \cite{Massar:1999wg},
they have  used an observation by Carlip and Teitelboim \cite{Carlip:1993sa}
that, if we consider a coupled system of a black hole and matter
exchanging energy between them, one needs to add a boundary term
to the bulk action,
$ S = S_{bulk} + A \Theta /8\pi G.$
Here, $\Theta =\kappa t$ for on-shell and stationary metrics.
Then in quantizing the system, 
the Wheeler-DeWitt equation $\hat{H}_{tot} \Psi =0$ in the bulk
must be supplemented by the boundary Schr$\ddot{\text{o}}$edinger equation \cite{Carlip:1993sa}
$ i\hbar \partial_{\Theta} \Psi = - (\hat{A}/8\pi G) \Psi, $
and the total system's wave function 
evolves as $\exp (iA\Theta/8\pi G \hbar)$ in the WKB approximation. 
\vspace{0.5em}
\\ 
Here, we argue that 
the ratio (\ref{ratio}) should be valid also for processes including
classical absorption of energy into the black hole, if we
generalize the notion of transition probabilities in the following way. 
The matter system outside the horizon dissipates the energy 
by transferring it into the black hole. 
Furthermore, the system feels a kind of thermal noise due to the 
Hawking radiation. Hence, by including both effects of the heat
transfer, in and out, at the horizon, the effective equation of motion 
for matter is controlled by a stochastic Langevin equation with 
dissipation and noise terms. 
In such a situation, one can define a probability distribution 
of the system to take some configuration. 
Time evolution of the probability distribution function 
is described by the Fokker-Planck equation. 
Clearly the time reversal symmetry is violated, and we can expect
an asymmetry between the probabilities of the forward and the reversed
processes. The ratio is  evaluated in general Langevin 
processes in \cite{Hatano:2001}. 
By applying it to our case, 
the energy transfer 
into the black hole can be rewritten as the area change of the black hole
through the first law of black hole thermodynamics
($\Delta S_{BH} =\Delta E/T_{BH}$). 
It can be possible to bring it together with Hawking radiation effect into the total amount of area change.
This suggests the validity of the probability ratio (\ref{ratio})
for wider situations including classical absorption of energy into
the black hole.
Details will be discussed in a forthcoming paper.

\vspace{1em}
{\it Non-equilibrium Fluctuations of Horizons.} --- We consider a 
sequence of configurations of a coupled system of a black hole and matter, and
denote it as $\Gamma =\{ {\cal C}_0(A_0), {\cal C}_1(A_1), \ldots , 
{\cal C}_M(A_M) \}$. The configuration ${\cal C}_k$
is realized at a discretized time $t=k \Delta t.$ 
Each transition probability is 
given by ${\cal W}_{\lambda^\text{F}(t_k)} ({\cal C}_k(A_k) \rightarrow {\cal C}_{k+1}(A_{k+1}))$.
If we assume the Markov process,
the transition probability for the sequence of configurations $\Gamma$ to be realized
is given by a product of them,
\begin{align}
P^\text{F}(\Gamma)=\prod_{k=0}^{M-1} {\cal W}_{\lambda^\text{F}(t_k)}
({\cal C}_k(A_k) \rightarrow {\cal C}_{k+1}(A_{k+1})).
\end{align}
As we argued, the sequence can represent a general process of 
absorbing and emitting matter through the black hole horizon.
On the other hand, the probability for the reversed sequence of configurations
$\Gamma^*=\{ {\cal C}_M(A_M), \ldots , {\cal C}_1(A_1), {\cal C}_0(A_0) \}$ 
with the reserved change of the external parameter is given by 
\begin{align}
P^\text{R} (\Gamma^*) &= \prod_{k=0}^{M-1} {\cal W}_{\lambda^\text{R}(t_k)}
( {\cal C}_{_{M-k}}(A_{_{M-k}})
\rightarrow {\cal C}_{_{M-k+1}}(A_{_{M-k+1}} ))
\nonumber \\
&=
\prod_{k=0}^{M-1} e^{\frac{\Delta A}{4G\hbar}}{\cal W}_{\lambda^\text{F}(t_k)}( {\cal C}_{k+1}(A_{k+1})
\rightarrow {\cal C}_k(A_k) )). 
\end{align}
Here we have used eq. (\ref{ratio}).
Hence the ratio of these two probabilities is given by 
\begin{align}
\frac{P^\text{F}(\Gamma)}{P^\text{R}(\Gamma^*)} = \exp \left(\frac{A_M -A_0}{4G \hbar}\right)
\equiv \exp (S_P(\Gamma)). 
\label{pathratio}
\end{align}
$S_P(\Gamma)$ is defined as the logarithm of the ratio, and
proportional to the difference of area,
which is not  necessarily small.
\vspace{0.5em}
\\ 
We now derive a Crook's type fluctuation theorem in the black hole system. 
The matter is assumed to be in thermal equilibrium with
Hawking temperature $T_{H}$ with an external parameter 
$\lambda^F (0)$ 
at the beginning.  First define the total dissipation $ \Delta S(\Gamma)$ by
\begin{align}
\exp (-\Delta S(\Gamma)) \equiv \frac{p_{\lambda^\text{R}(0)} ({\cal C}_M)}{p_{\lambda^\text{F}(0)} ({\cal C}_0)} \exp (-S_P(\Gamma)),
\label{SgammaDef}
\end{align}
where $p_{\lambda^{(\text{F, R})}(t_0)}$ is the initial probability distribution for matters under the forward or reversed protocols.
We assume that these probability distributions are canonical distributions with the Hawking temperature.
A relation $\Delta S(\Gamma^*) = - \Delta S(\Gamma)$ is followed by
$S_P(\Gamma^*) = - S_P(\Gamma)$.
A transition probability to produce the total dissipation $\Delta S(\Gamma)=\Delta S$
under the forward protocol $\lambda^\text{F}(t)$ is now given by 
\begin{align}
\rho^\text{F}(\Delta S) =\sum_{{\cal C}_0,\Gamma}  p_{\lambda^\text{F}(0)} ({\cal C}_0)
P^\text{F}(\Gamma) \delta (\Delta S(\Gamma) -\Delta S)). 
\end{align}
Here, $\sum_{{\cal C}_0} p_{\lambda^\text{F}(0)} ({\cal C}_0)$ stands for 
a sum over all possible initial states weighted by the initial distribution.
$\sum_{\Gamma}$ is a path integral for all possible trajectories.
By using (\ref{pathratio}) and (\ref{SgammaDef}), it is straightforward to
show the Crook's type fluctuation theorem
\begin{align}
\rho^\text{F} (\Delta S) e^{-\Delta S} = \ \rho^\text{R}(-\Delta S).
\label{CrooksBH}
\end{align}
From the definition of $\Delta S(\Gamma)$ in (\ref{SgammaDef}),
it is identified as a sum of the black hole entropy and the entropy of matter;
\begin{align}
\Delta S(\Gamma) = \frac{\Delta A}{4G\hbar} + \beta_{H} (\Delta E - \Delta F). 
\end{align}
By integrating the equation (\ref{CrooksBH}) over $\Delta S$,
it gives a Jarzynski type equality 
\begin{align}
\langle e^{-\Delta S} \rangle =1. 
\label{JarBH}
\end{align}
The Jensen inequality implies the generalized second law of 
thermodynamics $\langle \Delta S \rangle \ge 0$ of the black hole with matter \cite{Bekenstein:1974ax}.
An important point here is that it is satisfied only in an averaged
sense, and in order to satisfy the Jarzynski type equality (\ref{JarBH}),
entropy decreasing processes must exist as individual processes,
and their probabilities are miraculously arranged to satisfy
the Jarzynski equality.
\\
\\
{\it Conclusions and Discussions.} --- In this letter, we have applied 
the recent developments in non-equilibrium statistical physics to 
area changing processes of a black hole interacting with external matter.
The non-equilibrium fluctuation theorems 
$\grave{a}$ la Crooks and Jarzynski
are derived, which lead to the generalized second law of black holes. 
The second law holds only after
taking a thermodynamic average, and it should be violated as an 
individual process in a miraculous way to satisfy the Jarzynski equality.
\vspace{0.5em}
\\
There are several issues that further investigations are necessary.
First, we have applied the formula (\ref{ratio}) to general processes
including absorption and emission of classical and quantum particles.
As argued, it will be proved by using a path integral formulation of
stochastic processes for the matter fields in a black hole background.
More details will be reported in a separate paper.
Instead of using the stochastic equation for the matter that
may be obtained after integrating out internal degrees of 
freedom of a black hole, we may consider a whole quantum
system containing both of the internal and external degrees of freedom.
This will be achieved by considering the global Hilbert space
of the left and right wedges altogether.
\vspace{0.5em}
\\
Another issue is to apply a different type of fluctuation theorems,
such as the steady state fluctuation theorem.
If the matter continues to be in a thermal equilibrium with a 
different temperature from $T_{H}$, there is a constant flow
of energy between the black hole and the matter.
The fluctuation theorem is also applicable to such a situation.
From the steady state fluctuation theorem,
we can derive a fluctuation-dissipation relation 
by assuming  Gaussian fluctuations 
around a long-time average in a late time. 
We may also be able to obtain a Green-Kubo relation that relates the correlation of energy flow and a
proportionality coefficient between area change of horizon and energy flow.
\vspace{0.5em}
\\
More ambitiously, as Einstein asked himself how to prove the reality of atoms, 
we may also ask ourselves, what is an analog of the   
Avogadro constant in the gravity,  or how we can probe such 
quantities by looking at some kinds of fluctuations caused by
the microscopic structures of the space-time. 
An analogous quantity to the Avogadro constant will be the black hole
entropy per unit area $\sim 10^{70}[1/m^2]$  (or its exponential). 
Since it is a huge number, the fluctuation will be largely suppressed. 
\\
Finally, note that the fluctuation of the black hole entropy (horizon area) is proportional to the
specific heat, which is negative for the Schwarzschild black hole.
As it means an instability of the black hole, it becomes the more
important to take into account non-equilibrium effects of the fluctuations.

\begin{acknowledgments}
The work was presented by one of us (S.I.) at the RIKEN workshop 
on 23 July 2010
and by S.O. at the YITP workshop YITP-W-10-02, Kyoto on 21-24 July 2010.
We would like to thank the participants for useful discussions and comments,
especially, K. Fujikawa, T. Izuyama and H. Kawai. 
We also acknowledge T. Sagawa for enlightening discussions 
on non-equilibrium physics.
The research by S.I. is supported in part by the Grant-in-Aid for Scientific Research (19540316) from MEXT, Japan.
The research by  S.Z. is supported in part by 
the Japan Society for the Promotion of Science Research Fellowship for Young Scientists. 
We are also supported in part by " the Center for the Promotion of Integrated Sciences (CPIS) "  of Sokendai.
\end{acknowledgments}

\bibliography{BHFT.bib}

\end{document}